\documentclass[aps,epsfig]{revtex4}   
\usepackage{graphicx}
\begin{document}
\def\be{\begin{equation}}
\def\ee{\end{equation}}
\def\bfi{\begin{figure}}
\def\efi{\end{figure}}
\def\bea{\begin{eqnarray}}
\def\eea{\end{eqnarray}}

\title{Scaling and universality in the aging kinetics of the two-dimensional clock model}

\author{Federico Corberi$^\dag$, Eugenio Lippiello$^\ddag$ 
and Marco Zannetti$^\S$}
\affiliation {Istituto Nazionale di Fisica della Materia, Unit\`a
di Salerno \\ and Dipartimento di Fisica ``E.Caianiello'', 
Universit\`a di Salerno,
84081 Baronissi (Salerno), Italy}

\dag corberi@na.infn.it \ddag lippiello@sa.infn.it 

\S zannetti@na.infn.it


\begin{abstract}
We study numerically the aging dynamics of the two-dimensional $p$-state clock model after a quench
from an infinite temperature to the ferromagnetic phase or to the Kosterlitz-Thouless phase. 
The system exhibits
the general scaling behavior characteristic of non-disordered coarsening systems.
For quenches to the ferromagnetic phase, the value of the dynamical exponents, 
suggests that the model belongs to the Ising-type 
universality class. Specifically, for the integrated response function 
$\chi (t,s)\simeq s^{-a_\chi}f(t/s)$, we find $a_\chi $ consistent with
the value $a_\chi =0.28$ found in the two-dimensional Ising model.
\end{abstract}

\maketitle

\section{Introduction}

The phase-ordering kinetics of systems quenched from an high-temperature
disordered state to an ordered phase or to a critical phase 
is characterized by the growth of a characteristic length $L(t)$.
In the late stage a power-law behavior
\be
L(t)\sim t^{\frac{1}{z}} 
\label{0intr}
\ee
sets in, where $z$ is the dynamical exponent, and dynamical scaling~\cite{Bray94} is observed.
Accordingly, configurations of the system at two subsequent times are 
statistically equivalent if lengths are measured in units of $L(t)$, namely
if the rescaled length $x=r/L(t)$ is considered.
This property is reflected by the analytical form of physical observables.
In the quench to a critical point, for example, the equal-time order parameter 
correlation function obeys the scaling form
\be
G(\vec r,t)\sim r^{-(d-2+\eta ) }g(x),
\label{firstintr}
\ee 
where $d$ is the spatial dimensionality and $\eta $ is the usual exponent
of static critical phenomena.    
Scaling behaviors such as Eqs.~(\ref{0intr},\ref{firstintr})
are generic for growth kinetics in non-frustrated, non-disordered systems and are
observed regardless of different specific details.
Concerning the values of the exponents, such as $z$ or others entering
different quantities, 
they are expected to take the same value for systems belonging to the
same non-equilibrium universality class.
It is well known that systems undergoing a second order equilibrium phase transition 
can be classified into static equilibrium universality classes according to the value of their 
critical indices. These are found to depend only on a small set of parameters,
such as space dimensionality or the number of components of the order parameter . 
In the same way, dynamic universality classes can also be introduced on the
basis of the value of dynamic exponents. This subject has been thoroughly
studied for the equilibrium critical dynamics~\cite{Hohenberg77}, where
the renormalization group provides the basic mechanism
for scaling and universality, analogously to the static case.   
By contrast, the same subject is not well understood in far from equilibrium
systems~\cite{Chatelain04,Odor04}.

In this paper we consider the phase-ordering kinetics of the clock model with a 
$p$-fold degenerate ground state in two dimensions. 
For $p\leq 4$ this model has a single second order phase transition
separating a disordered from an ordered phase.
For $p\geq 5$ a Kosterlitz-Thouless (KT) critical phase also exists
and the phase transitions are of the Kosterlitz-Thouless
type. 
We study numerically the model with $p=3$ and $p=6$ and a non-conserved order parameter, 
quenched to the ordered region
and, for the case $p=6$, also to the KT phase.
In previous works~\cite{Kaski83,Liu93} the growth law~(\ref{0intr}) and the scaling 
of $G(\vec r,t)$ in quenches to the ordered region was 
analyzed; for quenches to a critical point, 
two time quantities were studied in~\cite{Chatelain04}. Here we
extend these results presenting a global analysis of the scaling properties
for quenches in the ordered and in the critical region, by considering 
one-time quantities, such as $G(\vec r,t)$, and two-time quantities, 
such as the autocorrelation function and the integrated response function.
We find that the scaling forms expected for all these quantities in non-disordered systems 
are obeyed,
although preasymptotic corrections are observed in the simulated range of times.
We may then conclude that the clock model exhibits the generic scaling behavior
characteristic of phase-ordering systems in the late stage of the dynamics.
This calls for the question of the value of the dynamical exponents and the issue of their
universality, namely whether the dynamics of the model 
is regulated by the same exponents of other coarsening systems and whether their value depends on $p$. 
Since in the KT phase critical exponents depend continuously on temperature, 
this problem is pertinent to the quench to the ordered phase.
Among the most usually considered two-dimensional statistical models of ferromagnetism, the Ising
and the XY model are those to which the clock model can be naturally compared.
These models are particular cases, with
$p=2$ and $p=\infty $, of the clock model. The former has a scalar order parameter
and a discrete symmetry. In the latter, the order parameter  
is a vector with $N=2$ components and there is a continuous $O(2)$ symmetry.
The symmetry group of the model, 
together with the spatial dimensionality, determines the types of 
topological defects.
Consequently, topological defects are interfaces or two-dimensional vortices in the
Ising and XY model, respectively. 
Since the nature of the topological defects controls several equilibrium properties 
and the late stage ordering~\cite{Bray94} these models belong to different universality classes both in
equilibrium and out of equilibrium. The clock model is, in some sense, intermediate
between the Ising and XY model, because the order parameter is a vector with two components, 
as in the XY model, but there is a finite degeneracy of the ground state, namely a discrete symmetry.
Topological defects are then both interfaces and vortices. 

In this paper we show that the exponents measured for the phase-ordering kinetics of the clock model 
with $p=3,6$ are consistent with those 
of the two-dimensional Ising model quenched below the critical
temperature. This suggests that systems with a finite
degeneracy of the ground state may belong to the Ising non-equilibrium universality class,
and that the presence of other topological defects besides interfaces does not affect
the universal properties of these systems.

This Article is organized as follows: In Sec.~\ref{model} we introduce the model
and define the main observables. In Sec.~\ref{ferromagnetic} we discuss the general
scaling behavior of systems quenched into an ordered region, present the
results of numerical simulations of the clock model with $p=3$ and $p=6$, and
compare them with the behavior of the Ising model and of the XY model.
In Sec.~\ref{kt} we discuss the scaling properties of coarsening systems quenched
to a critical phase and the results of numerical simulations of the clock model
with $p=6$ quenched into the KT phase. Sec.~\ref{concl} contains the final observations
and the conclusions.

\section{Model and observables} \label{model}

The $p$-state clock model is defined by the Hamiltonian
\be
H[\sigma ]=-J\sum _{<ij>}\vec \sigma _i \cdot \vec \sigma _j=
-J\sum _{<ij>}cos(\theta _i-\theta _j),
\label{hamiltonian}
\ee
where $\vec \sigma _i$ is a two-components unit vector spin pointing along  
one of the directions
$\theta _i=2\pi n_i/p$, with $n_i \in \{1,2,...,p\}$, and $<ij>$
denotes nearest neighbors sites $i,j$ on a lattice.
We will consider
a square lattice in spatial dimension $d=2$. This spin system is 
equivalent to the Ising model for $p=2$ and to the XY model 
for $p\to \infty $. 

For $p\leq 4$ the clock model has a critical point 
separating a disordered from an ordered phase at $T=T_1$. 
For $p\geq 5$ there exist two transition 
temperatures $T_1$ and $T_2>T_1$~\cite{jose}. For $T<T_1$ the
system is ferromagnetic, and for $T>T_2$ it is in a paramagnetic
phase. Between these two temperatures, for $T_1<T<T_2$, 
a KT phase~\cite{kt} exists where the correlation function
behaves as $G_{eq}(r)\sim \mid r\mid ^{-\eta(T)}$ with the anomalous
dimension $\eta (T)$ continuously depending on the temperature.
Both the transitions are of the KT type, namely the correlation length 
diverges exponentially as $T_1$ or $T_2$ are approached from
the ferromagnetic or paramagnetic phase, respectively.
Approximate analytic results~\cite{jose} predict 
\be
T_1/J=\frac {4\pi ^2}{1.7 p^2}
\label{T2}
\ee
and $T_2$ to coincide with the KT temperature of the XY model
$T_2/J=0.95$ (here and in the following we set the Boltzmann constant
$k_B=1$). The exponent $\eta (T)$ is expected~\cite{jose} to vary
between $\eta (T_1)=4/p^2$ and $\eta (T_2)=1/4$.
Numerical simulations~\cite{Challa86} are consistent with these 
predictions. 

A dynamics is introduced by randomly choosing a single spin 
and updating it with Metropolis transition rate
\be
w([\sigma]\to [\sigma'])=\min \left [1, \exp (-\Delta E/T)\right ].
\label{metropolis} 
\ee
Here $[\sigma]$ and $[\sigma']$ are the spin configurations before and
after the move, and $\Delta E=H[\sigma']-H[\sigma]$.

We consider the protocol where the system is initially prepared in an
high temperature uncorrelated state and then quenched, at time $t=0$,
to a final temperature $T_f$ in the ferromagnetic phase or in the KT phase.
The characteristic size $L(t)$ grows until it becomes
comparable with the system size and the new equilibrium state
at $T_f$ is globally attained. For an infinite system the
final equilibrium state is never reached and $L(t)$ keeps growing
indefinitely. In the late stage the power law~(\ref{0intr}) sets in. 
The characteristic length $L(t)$ can be estimated from the
knowledge of the two-points equal time correlation function
\be
G(r,t)=\langle \vec \sigma_i(t) \cdot \vec \sigma_j(t) \rangle
\label{gdir}
\ee
where $\vec \sigma_i(t)$,$\vec \sigma_j(t)$ are spin variables at time $t$
on two lattice sites whose distance is $r$, and $\langle \dots \rangle$
means an ensemble average, namely taken over different initial conditions
and thermal histories.
Due to homogeneity, $G(r,t)$ does
not depend separately on $i$ an $j$ but only on $r$.
Enforcing this property we will numerically compute the correlation 
in the following as
\be
G(r,t)=\frac {1}{4N}\sum _i\sum _{j\in J_+ }
\langle \vec \sigma_i(t) \cdot \vec \sigma_j(t) \rangle,
\label{numgdir}
\ee
where $i$ runs over all the $N$ sites of the lattice and $J_+$
is the set of four points reached moving a distance $r$ from $i$
along the horizontal or vertical directions. 
The methods to extract $L(t)$ from $G(r,t)$ depend on the scaling properties
of $G(r,t)$ and differ if quenches in the ferromagnetic or
in the KT phase are considered, as discussed in 
Secs.~\ref{ferromagnetic},\ref{kt}.

The two time quantities that will be considered in this paper
are the autocorrelation function
\be
C(t,s)=\langle \vec \sigma_i (t)\cdot \vec \sigma_i (s) \rangle
\label{autocorr}
\ee
and the integrated (auto)response function, or zero field cooled susceptibility  
\be
\chi (t,s)=\int _s ^t dt' R(t,t').
\label{integrated}
\ee
The quantity 
\be
R(t,t')=\sum _{\alpha} \left . \frac {\partial \langle \sigma_i^\alpha (t) \rangle}
{\partial h_i^\alpha (t')}\right \vert _{h=0},
\ee
$\alpha $ being a generic vector component, is the response function associated 
to the perturbation caused by an impulsive magnetic field $\vec h_i$
switched on at time $t'<t$.
Recently, new efficient methods for measuring the response function without applying 
the perturbation have been introduced~\cite{Chatelain03,Ricci03,Lippiello05}.
In the following we will use the one derived in~\cite{Lippiello05}.
For spin systems subjected to a Markovian dynamics
an out of equilibrium generalization
of the fluctuation dissipation theorem was derived~\cite{nota}, 
relating the response functions
to particular correlation functions of the unperturbed system.
For the integrated response function~(\ref{integrated}) it reads
\be
T\chi (t,s)=\frac {1}{2}\left [C(t,t)-C(t,s)-
\int _s ^t \langle \vec \sigma_i(t)\cdot \vec B_i(t')\rangle dt' \right ].
\label{algochi}
\ee
where
\be 
\vec B_i[\sigma]=-\sum _{\vec \sigma '}(\vec \sigma_i-\vec \sigma_i') w([\sigma]\to [\sigma']).
\label{b}
\ee
In this equation $[\sigma]$ and $[\sigma']$ are two configurations differing only by
the spin on site $i$, taking the values $\vec \sigma_i$ and $\vec \sigma_i'$ respectively.   
The notation $\langle \vec \sigma_i(t)\cdot \vec B_i(t')\rangle$ 
in Eq.~(\ref{algochi}) means the average  
$\sum _{[\sigma],[\sigma']}\vec \sigma_i\cdot \vec B_i [\sigma ']
p([\sigma],t;[\sigma'],t')p([\sigma'],t')$, where 
$p([\sigma'],t')$ is the probability to find the configuration
$[\sigma ']$ at time $t'$
and $p([\sigma],t;[\sigma'],t')$ is the joint probability
between $[\sigma]$ at time $t$ and $[\sigma']$ at time $t'$.  

Eq.~(\ref{algochi}) allows to compute the integrated response function by
measuring correlation functions on the unperturbed system, avoiding the
complications of the traditional methods where a perturbation is
applied, and improving significantly the quality of the results~\cite{Lippiello05}. 

\section{Quenches to $T_f<T_1$}\label{ferromagnetic}

\subsection{General scaling properties}\label{ferromagnetic1}

Let us start considering quenches from an high temperature 
disordered phase to the ferromagnetic region.
In this case one observes the growth of compact domains separated by
topological defects such as interfaces or vortices
(see Fig.~\ref{figdomainsp3}).
As a consequence, a sharp distinction can be made in the late stage between spins 
belonging to the interior of domains from those pertaining to
the defects. The interior of the domains very soon attains 
local equilibration in one of the broken symmetry equilibrium
phases at $T_f$, whereas the degrees of freedom around defects are out of equilibrium and
are responsible for the aging of the system. Observables such as 
$C(t,s)$, $\chi (t,s)$ and $G(r,t)$ take an addictive structure~\cite{Corberiteff}
\be
C(t,s)=C_{st}(t-s)+C_{ag}(t,s),
\label{addictive}
\ee
and similarly for $\chi (t,s)$ and $G(r,t)$. In the following, for clarity, we will 
focus on $C(t,s)$, but similar considerations hold for the other quantities. 
The dynamics of the spins in the bulk of domains
provide the equilibrium contribution $C_{st}(t-s)$ while what is left over,
$C_{ag}(t,s)$, accounts for the aging behavior. 
Since equilibrium dynamics is well understood, the behavior of $C_{st}(t-s)$
is generally well known. In particular, at $T_f=0$
equilibrium dynamics is frozen and $C_{st}(t-s)\equiv 0$.  
On the other hand, much interest is focused on the aging part of the aforementioned
observables, which is less understood. 
This contribution can be isolated by subtracting $C_{st}(t-s)$, computed
in equilibrium, from $C(t,s)$. 
However, for models with a discrete symmetry, it is computationally much more efficient
to resort to a different method. This amounts to study a modified system where 
$T_f$ in the transition rate~(\ref{metropolis})
is set equal to zero if the spin $\vec \sigma _i$ to be updated
belongs to the bulk, namely if it is aligned with all its neighbors. 
Since the bulk degrees of freedom, which alone contribute to $C_{st}(t-s)$,
feel $T_f=0$, and $C_{st}(t-s)\equiv 0$ at $T_f=0$,
by computing observables with this modified dynamics one isolates
the aging term leaving other properties of the dynamics 
unchanged~\cite{Corberiresp2d}. In order to check this we have
computed $C(t,s)$ and $C_{st}(t-s)$ with the Glauber dynamics
in a system with $p=3$ quenched to $T_f=1<T_1$ and in an equilibrium 
system at the same temperature. In Fig.~(\ref{figcadded}) we compare 
$C_{ag}(t,s)$ obtained by subtraction of these quantities through Eq.~(\ref{addictive}) 
(symbols) and the same quantity obtained directly by means of a 
quench simulation with the modified dynamics (continuous lines). 
This figure shows an excellent agreement,
confirming that the modified dynamics is efficient and accurate.  
In the remaining of this Section, therefore,
we will always present results obtained with this modified dynamics.
\begin{figure}
    \centering
    \caption{\\(Color online) $C_{ag}(t,s)$ obtained with the two methods described in the text
             is plotted against $t-s$ for $s=100,200,400,800,1600,3200$.}
\label{figcadded}
\end{figure}

In the late stage of the evolution, after a characteristic
time $t_{sc}$ when $L(t)$ is much larger than all other microscopic
lengths, dynamical scaling is obeyed~\cite{Bray94,Liu93}. Accordingly, for the
correlation function one has
\be
G_{ag}(r,t)=M^2 g(x),
\label{scalgferro}
\ee  
where $x=r/L(t)$ and $\vec M$ is the equilibrium magnetization at $T_f$.  
In systems with a discrete symmetry and sharp interfaces, a short distance behavior
($x\ll 1$) of the type $1-g(x)\sim x$ is found~\cite{Bray94,Liu93}, namely a Porod's
tail $\hat G(\vec k,t)\sim k^{-(d+1)}$ in momentum space for
large $k$ ($k\gg L(t)^{-1}$). This is known to be true, in particular, for
the clock model, for all $p<\infty $, although the whole form of $g(x)$ 
depends on $p$~\cite{Liu93}. In systems with a vector order parameter and an
$O(N)$ symmetry one has a generalization of the Porod's law~\cite{Mondello90},
 $\hat G(\vec k,t)\sim k^{-(d+N)}$. 

From Eq.~(\ref{scalgferro}) one can extract a quantity $L_1(t)$
proportional to the typical domain size $L(t)$ from the condition
$g(x)=\frac{1}{2}$,
namely as the half-height width of $G_{ag}(r,t)$.
Alternatively, a characteristic length $L_2(t)$ can be
extracted as
$L_2(t)=\int dr \, G_{ag}(r,t)$.
Clearly, if Eq.~(\ref{scalgferro}) holds, $L_1(t)\propto L_2(t)$.

The size of domains can be related to the
density of defects $\rho (t)$.
For a coarsening system where topological defects are only interfaces, such
as the Ising model, 
one has a power law behavior $\rho (t)\propto t^{-\delta}$, with~\cite{Bray94} 
\be
\delta=1/z.
\label{deltascal}
\ee
This result does not apply to systems with different defects, such as vortices or others. 
In the case of vector $O(N)$ (with $N\ge 2$) 
model~\cite{Bray94,Blundell94} one has 
\be
\delta=2/z,
\label{deltavec}
\ee
and logarithmic corrections for $N=2$.
In these cases $\rho (t)$ provides an indirect, 
alternative method for the determination of $L(t)$, and hence of $z$.
For the clock model, where defects are interfaces and vortices, neither 
Eq.~(\ref{deltascal}) or Eq.~(\ref{deltavec}) can be straightforwardly used.
However, a simple inspection of the configurations 
(see Fig.~\ref{figdomainsp3}) suggests that,
since vortices are point like, their contribution to $\rho (t)$
must be negligible in the late stage.
Therefore we expect Eq.~(\ref{deltascal}) to be obeyed asymptotically.
  
The dynamical exponent $z$ is believed to be universal
for quenches to the ordered phase $T_f< T_1$:  
the same value $z=2$ as for the Ising model is 
expected for every value of $p$~\cite{Kaski83,Liu93} in the clock model and
for every $N$ in $O(N)$ models. 

Coming to two-times quantities, the aging part of the autocorrelation function
is expected~\cite{Bray94,Corberiteff} to scale as
\be
C_{ag}(t,s)=h(y),
\label{scalcferro}
\ee
with $y=t/s$ and $h(y)\sim y^{-\lambda/z}$ for $y\gg 1$. The exponent $\lambda $ 
is believed to be the Fisher-Huse exponent which regulates the large $t$ decay of
the initial condition autocorrelation function $C(t,0)\sim t^{-\lambda /z}$.
In the Ising model one has $\lambda = 5/4$.
We are not aware of a systematic study of this exponent in the XY model.
In~\cite{Berthier01} it is argued that this exponent depends on
$T_f$ and, for the particular case $T_f=0.3$ the value $\lambda=0.54$ is found.
For the integrated response function scaling implies 
\be 
\chi _{ag}(t,s)=s^{-a_\chi }f(y). 
\label{scalchiferro}
\ee
For $p=2$ the scaling function behaves as
\be
f(y)\sim y^{-a_{\chi}},
\label{altromodo}
\ee
for $y\gg 1$. 
Regarding the exponent $a_{\chi}$, analytical calculations in solvable 
scalar models or in the large-$N$ model~\cite{altri1,ninf}
find the following dependence on dimensionality
\be
    a_{\chi} = \left \{ \begin{array}{ll}
        \delta \frac{d-d_L}{d_U-d_L}  \qquad $for$ \qquad d < d_U  \\
        \delta  \qquad $with log corrections for$ \qquad d=d_U \\
	\delta   \qquad $for$ \qquad d > d_U, 
        \end{array}
        \right .
        \label{aphenom}
        \ee
where $d_L$ is the lower critical dimension of static critical phenomena and
$d_U$ is an upper dimension that turns out to be $d_U=3$ or $d_U=4$ for systems
with a discrete or continuous symmetry. This expression shows that the response of coarsening
systems depends on dimensionality in a non trivial way. 
Numerical simulations~\cite{altri1,altri2,claudio} of scalar and
 ${\cal O}(N)$ vectorial systems, with conserved and non conserved order parameter, are consistent
with Eq.~(\ref{aphenom}). 
The value of $a_\chi $ has never been investigated for systems
with a discrete symmetry and a degeneracy of the ground state larger than $p=2$, as
in the case considered in this paper. 

Notice that, for a given dimensionality $d<d_U$, Eq.~(\ref{aphenom}) predicts a different 
exponent for systems with a continuous or a discrete symmetry.
In the case $d=2$, for instance,
for the Ising model Eq.~(\ref{aphenom}) gives $a_{\chi }=1/4$ while for the XY model 
$a_\chi =0$. Therefore, for the model under investigation, the value of $a_\chi$ may be
used to discriminate between the Ising and XY non-equilibrium universality classes.

Finally, let us recall that the scaling 
behaviors~(\ref{scalgferro},\ref{scalcferro},\ref{scalchiferro}) 
are only expected asymptotically. 
Since numerical simulations can only access a finite time region, preasymptotic effects
may be present. In particular in numerical simulations of the
Ising model with a non-conserved order parameter, one usually observes an effective exponent 
$1/z_{eff}\simeq 0.48$ in place of $1/z=0.5$. 
The integrated response function has been
also shown~\cite{Chatelain03,Corberiresp2d} to be affected by corrections 
to scaling. 
These can be conveniently discussed in terms of the effective exponent, defined as
\be
a_{eff}(y,s)=- \left . \frac{\partial \ln \chi _{ag}(t,s)}{\partial \ln s} \right \vert _y. 
\label{aeff}
\ee
With a scaling form such as~(\ref{scalchiferro}),
one would have $a_{eff}(y,s)=a_\chi$, independent of $y$ and $s$. 
However, if preasymptotic effects are present, the
effective exponent takes a value which depends both
on $y$ and $s$. For $p=2$ it was shown~\cite{Corberiresp2d} that, because of this,
$a_{eff}(y,s)$ is found in numerical simulations in the range
$0.25\le a_{eff}(y,s)\le 0.28$.

\subsection{Numerical results}\label{ferromagnetic2}

In the following we will present the numerical results. 
Setting $J=1$, for each case considered we simulated a square lattice of size
$1000^2$ with periodic boundary conditions and
an average over 100 realizations was performed. 
Statistical errors, when not explicitly plotted in the figures, are
comparable to the thickness of the lines. 

For $p=3$ there is a ferro-paramagnetic transition at $T_1\simeq 1.326$
while for $p=6$, according to Eq.~(\ref{T2}), one has $T_1\simeq 0.645$.
We performed a series of simulations of quenches to $T_f<T_1$, 
with $T_f=1/2$ or $T_f=1$ for $p=3$ and $p=6$, respectively. 
Typical configurations of domains in the late stage are shown in Fig.~\ref{figdomainsp3}. 
Notice the simultaneous presence of interfaces and vortices. These are defined analogously
to those of $O(N)$ models: on encircling a vortex the order parameter rotates
by $\pm 2\pi $ (although in the clock model rotations are obtained by
discrete steps). While for $p=3$ vortices and interfaces between
different phases are all energetically equivalent, 
in the case $p=6$ one has the additional feature of 
different kinds of vortices and interfaces. 
Let us consider a domain characterized by having all the spins
pointing along the direction $\theta =2\pi n/p$. Spins belonging to the domain are 
characterized by having the same value of $n_i$, $n_i=n$. This domain can be separated by an
interface from another domain where spins point along a different direction 
$\theta =2\pi m/p$. Clearly, interfaces between domains of contiguous phases, 
namely with $n=m\pm 1$ are energetically less expensive then the others.
The more energetically expensive interfaces are eliminated faster from 
the system (they are already
practically absent in Fig.~\ref{figdomainsp3}).
This fact influences considerably the topology of the growing structure.
For instance, one clearly observes that between two domains of
non contiguous phases, say with $n=m$ and $n=m+2$, a thin slab of phase $n=m+1$
is interposed in order to minimize energy. 

Analogously one observes also different kinds of vortices. Points 
where six phases meet are energetically favored, but vortices where a
lower number of phases meet can also be present. 
Moreover there are also points where
four (or more) domains meet, but two of them belong to the same phase: 
Encircling such points
one may enter domains characterized respectively by, say, the sequence $n$, 
$n+1$, $n+2$, $n+1$ again.
Clearly, encircling the most energetically favored vortices one finds
all the phases according to the sequence $n,n+1,n+2,n+3,n+4,n+5,n+6$
(or in reverse order). As for the interfaces, the high energy vortices
are quickly removed and a typical late stage configuration, as that of 
Fig.~\ref{figdomainsp3},
contains practically only the lowest energy vortices. 
The presence of all these different kind of defects in the system is possibly
the origin of the long lasting preasymptotic effects discussed below.

\vspace{2cm}
\begin{figure}
    \centering
   \hspace{1cm}
    \caption{(Color online) Configuration of the system with $p=3$ (left) and $p=6$ (right), at $t=3200$.}
\label{figdomainsp3}
\vspace{2cm}
\end{figure}
\newpage
A comparison between $L_1(t),L_2(t)$ and $\rho (t)^{-1}$ is shown in Fig.~\ref{figlengthp3}. 
After an early stage when domains are formed and scaling does not hold,
$L_1(t),L_2(t)$ and $\rho (t)^{-1}$ start growing with an approximate
power law behavior and for long times one has $L_1(t)\propto L_2(t)\propto \rho ^{-1}(t)$
(for $p=6$, $\rho (t)^{-1}$
does not obey a power law behavior in the range of simulated times. However, for the
longest simulated times
the effective exponent seems to approach a value roughly comparable
with that of $L_1(t)$ and $L_2(t)$).
This implies that Eq.~(\ref{deltascal}) is obeyed asymptotically. We recall that
for a system containing only vortices, such as the XY model, one would instead expect
the relation~(\ref{deltavec}).
Regarding the coarsening exponent, in the decade $10^4-10^5$
for $p=3$ we measure $1/z_{eff}=0.486\pm 0.002$, 
$1/z_{eff}=0.484\pm 0.002$ and $1/z_{eff}=0.478\pm 0.002$ from $L_1(t)$, 
$L_2(t)$ and $\rho (t)^{-1}$, while for $p=6$ we get 
$1/z_{eff}=0.467\pm 0.003$, $1/z_{eff}=0.474\pm 0.003$,  
$1/z_{eff}=0.450\pm 0.008$.
These values (apart from the last one which is evidently a still preasymptotic effective
exponent) are compatible with the value $1/z=1/2$ of the Ising model.
The different initial behaviors of
$L_1(t),L_2(t)$ and $\rho (t)^{-1}$, signal that
preasymptotic effects are present up to very long times. This 
is probably related to the presence of different types of defects.
Note also that, at the longest time considered, the density of defects
with $p=6$ is more than three times larger than with $p=3$.

\begin{figure}
    \centering
    \caption{(Color online) Comparison between $L_1(t),L_2(t)$ and $\rho (t)^{-1}$ for $p=3$ (left) 
             and $p=6$ (right).}
\label{figlengthp3}
\vspace{2cm}
\end{figure}

In Fig.\ref{figgrp3} we test the scaling form~(\ref{scalgferro}) of the equal time
correlation function. We plot $G_{ag}(r,t)/M^2$ against $x=r/L_1(t)$ for several values of
$t$ in the two decades range $[6.4\cdot 10^2-6.4\cdot 10^4]$. 
The data show a
good collapse on a single master-curve $g(x)$.
For small $x$ the Porod's behavior $1-g(x)\sim x$ is very neatly observed. 
\begin{figure}
    \centering
   \hspace{1cm}
    \caption{(Color online) Data collapse of $G_{ag}(r,t)$ against $x=r/L_1(t)$ for several times 
$t_n$ generated from $t_n=$Int$[\exp (n/2)+1]$ with 
$n$ ranging from 13 to 22 and $p=3$ (left) or $p=6$ (right).}  
\label{figgrp3}
\end{figure}

We turn now to consider two time quantities.
In Fig.\ref{figscalcp3} the autocorrelation function is plotted against $y$
for different values of $s$ in the range $[100-3200]$. 
As already observed when discussing Fig.~\ref{figlengthp3}, scaling is only approximatively 
obeyed in this regime. This is mirrored by $C_{ag}(t,s)$: Full data
collapse is not found. Regarding Fig.~\ref{figlengthp3},
we also noticed that scaling improves as time gets larger and it
is reasonably well obeyed for the longest simulated times.
The same conclusion can be drawn, for $p=3$, from $C_{ag}(t,s)$.
Indeed, in Fig.\ref{figscalcp3} one can observe that the data collapse improves
pushing $s$ and $y$ to larger values. 
In fact, although the data collapse of the curves is poor for small $y$
it gets better increasing $y$ and, for $y>10$, all the curves collapse. 
Moreover, the quality of the collapse improves also increasing $s$. 
Indeed, while the curves with small $s$ do not collapse (except, as anticipated, for
$y$ as large as $y>10$) the two curves with the largest values of $s$ 
($s=1600$ and $3200$) practically coincide for all $y>2$.
For $p=6$ the collapse is worse. 
Coming to the asymptotic behavior of the scaling function $h(y)\sim y^{-\lambda /z}$,
it is numerically too demanding to reach the asymptotic large-$y$ region
with the values of $s$ considered in Fig.~\ref{figscalcp3}. Then, we have 
evaluated $\lambda $ from the large $t$ behavior of $C(t,0)$. This quantity is
shown in the insets of Fig.~\ref{figscalcp3}. In the range $t\in [4\cdot 10^4, 10^5]$ we find
$\lambda /z=0.61 \pm 0.01$ and $\lambda /z=0.57 \pm 0.01$ for $p=3$ and $p=6$. 
For $p=3$ this value is consistent with the value $\lambda /z=5/8=0.625$
of the Ising model, keeping also into account that the effective exponent we
measure is still slightly increasing at the longest simulated times.
For $p=6$ the measured exponent is somewhat smaller than for $p=3$
and for the Ising model, but the fact that it keeps still growing at the longest
simulated times suggests that asymptotically the same value $\lambda =5/8=0.625$
could be obtained. This results suggest that there could be a unique non-equilibrium
universality class for every $2\le p<\infty$. 
\begin{figure}
    \centering
   \hspace{1cm}
    \caption{(Color online) $C(t,s)$ is plotted against $y$ for $p=3$ (left) with $s=100,200,400,800,1600,3200$ 
             and for $p=6$ (right) with $s=400,800,1600,3200$. 
             In the insets $C(t,0)$ is plotted against $t$.}
\label{figscalcp3}
\end{figure}
This hypothesis can be tested by considering the integrated response function.
This quantity, for $p=3$, is plotted against $t$ in Fig.~\ref{figchiaggiunta}, showing
a marked dependence on $s$.
Starting from zero at $t=s$, $\chi _{ag}(t,s)$  reaches a maximum at $t\simeq 2s$
and then decreases to zero with a power law behavior. A similar behavior is observed for $p=6$. 
According to Eq.~(\ref{scalchiferro}),
by fixing $y$ to a certain value and varying $s$ or equivalently $t$, 
the data should follow a power law with exponent
$-a_\chi$. As an example, the points corresponding to $y=4$, 
which have been marked with stars in the log-log plot of Fig.~(\ref{figchiaggiunta}), 
are approximatively aligned on a straight line of slope $0.26\pm 0.01$.
A similar analysis can be performed for every value of $y$.

\begin{figure}
    \centering
    \caption{(Color online) $T_f\chi _{ag}(t,s)$ is plotted against $t$ for fixed values of $s$.
     The dashed line is the power law $t^{-0.26}$.} 
\label{figchiaggiunta}
\vspace{2cm}
\end{figure}
In order to make a quantitative analysis of this exponent, 
and to detect preasymptotic effects, in Fig.~(\ref{figscalchi1p3})
we plot $\chi _{ag}(t,s)$ for fixed values of $y$ against $s$.
According to Eq.~(\ref{aeff}) the slope of
these lines is $a_{eff}(y,s)$ and, if scaling~(\ref{scalchiferro}) holds, 
one should find $a_{eff}(y,s)\equiv a_\chi$.
Preasymptotic effects, instead, introduce a weak dependence of this exponent
on $s$ and $y$. 
Apart from the curve $y=2$, which corresponds to very early times,
the slopes of all the curves are compatible with Eq.~(\ref{aphenom}), 
namely with $a_\chi=1/4$
(for $p=3$ in the range $s\in [800-3200]$ we find $a_{eff}(y,s)=0.25-0.30$, depending on $y$.
For $p=6$ the effect of preasymptoticity
is smaller and one finds $a_{eff}(y,s)\leq 0.27$ for every value of $y$ and $s$).
This pattern of behavior
of $a _{eff}(y,s)$ is similar to what is observed in the Ising model
where $a_{eff}(y,s)$ ranges in the interval, $a_{eff}(y,s)=[0.25-0.28]$. 

The data collapse of $s^{1/4} \chi _{ag}(t,s)$ vs $y$
expected from Eq.~(\ref{scalchiferro}) is shown in Fig.~\ref{figscalchi2p3}. 
For $p=6$ the collapse of the two curves
with the largest $s$ ($s=1600,3200$), is  
good at sufficiently large $y$ ($y>4$). It is 
poorer for $p=3$.
Note also that the asymptotic
behavior $f(y)\sim y^{-1/4}$ for $y\to \infty $ is well obeyed, consistently 
with Eq.~(\ref{altromodo}),
again confirming $a_\chi =1/4$ and ruling out the value $a_\chi =0$ appropriate to the XY model. 
These results 
strengthen the conclusion that 
the clock model below $T_1$ does not belong to the XY 
universality class and that the non equilibrium universality class is
the same for all $2\le p<\infty$.

\begin{figure}
    \centering
   \hspace{1cm}
    \caption{(Color online) $T_f\chi _{ag}(t,s)$ is plotted against $s$ for $p=3$ (left) and $p=6$ (right),
             with fixed values of $y$
    ($y=2,4,6,8,10,12,14,16,18$ from top to bottom). Numerical values are marked with
    error bars, continuous lines are guides for the eye.}
\label{figscalchi1p3}
\vspace{2cm}
\end{figure}

\begin{figure}
    \centering
   \hspace{1cm}
    \caption{(Color online) $T_fs^{1/4}\chi _{ag}(t,s)$ is plotted against $y$ for fixed values of $s$
      for $p=3$ (left) and $p=6$ (right).}
\label{figscalchi2p3}
\vspace{2cm}
\end{figure}


\section{$p=6$: Quenches to $T_1\leq T_f \leq T_2$} \label{kt}

Let us consider the model with $p=6$ quenched to the 
critical region $T_1\leq T_f \leq T_2$.
In the late stage the correlation function obeys
Eq.~(\ref{firstintr}).
From this equation one has
\be
I(t)=\int dr G(r,t) \propto L(t)^{3-d-\eta }\sim 
t^{\frac {3-d-\eta }{z}}. 
\label{intkt}
\ee
$L(t)$ can then be extracted as
\be
L(t)\propto I(t)^{\frac {1}{3-d-\eta }}.
\label{lung}
\ee

The autocorrelation function obeys~\cite{multip,Corberiteff} 
\be
C(t,s)= (t-s+t_0)^{-\frac {(d-2+\eta )}{z}}\tilde h(y),
\label{secintr}
\ee
where $t_0$ is a microscopic time.
Neglecting $t_0$ for $t-s\gg t_0$, $C(t,s)$ can be rewritten in scaling form
\be
C(t,s)\simeq s^{-\frac {(d-2+\eta )}{z}}h(y)
\label{secintr2}
\ee
where $h(y)=(y-1)^{-(d-2+\eta )/z}\tilde h(y)$, with the property
$h(y) \sim y^{-\lambda/z}$ for $y\gg 1$. Notice that, when using the $y$ variable 
in the large-$s$ limit, the condition $t-s\gg t_0$ for the validity of Eq.~(\ref{secintr2})
becomes $y>1$. For all $y>1$, then, one should expect data collapse of the curves $s^{(d-2+\eta )/z}
C(t,s)$ against
$y$ for different choices of $s$. 
 
The response function is given by~\cite{multip,Corberiteff}
\be
R(t,t')= (t-t'+t_0)^{-\frac{(d-3+\eta )}{z}}\tilde f \left (\frac {t'}{t}\right ).
\label{eerree}
\ee
Splitting the integral of Eq.~(\ref{integrated}) into two 
integration domains, introducing the arbitrary number $\epsilon $,
the integrated response function can be written as
\be
\chi (t,s)\sim t^{-\frac{(d-2+\eta )}{z}}\left [\int _y ^{1-\epsilon }du 
\,(1-u+\frac{t_0}{t})^{-\frac{(d-3+\eta )}{z}}\tilde f(u)
+\int _{1-\epsilon } ^1 du \,
(1-u+\frac{t_0}{t})^{-\frac{(d-3+\eta )}{z}}\tilde f(u)\right ]=
\label{interm}
\ee
\be
t^{-\frac{(d-2+\eta )}{z}}\left [I_1(y,\frac{t_0}{t},\epsilon)+I_2(\frac{t_0}{t},\epsilon)\right ],
\label{interm2}
\ee
where $u=t'/t$.
For large times $t$ one can choose $t_0/t \ll \epsilon \ll 1/2$. The condition
$t_0/t \ll \epsilon$ allows one to neglect $t_0/t$ with respect to $1-u$ in the first integral,
whose value can then be evaluated as $I_1(y,t_0/t,\epsilon )\simeq F(1-\epsilon,t_0/t)-F(y,0)$,
where $dF(u,t_0/t)/du=(1-u+\frac{t_0}{t})^{-\frac{(d-3+\eta )}{z}}\tilde f(u)$. 
Let us consider now the second integral $I_2(t_0/t,\epsilon )=F(1,t_0/t)-F(1-\epsilon,t_0/t)$.
Here, since $\epsilon \ll 1/2$, one can set $\tilde f (u)\simeq f(1)$, so that 
$F(1,t_0/t)\simeq (t_0/t) ^{-(d-2+\eta )/z}\tilde f(1)z/(d-2+\eta)$.
One then arrives at
\be
\chi (t,s)= t^{-\frac{(d-2+\eta )}{z}}f(y)+ 
\frac {\tilde f(1)}{\frac{d-2+\eta }{z}}t_0^{-\frac{d-2+\eta }{z}},
\label{intermm}
\ee
where $f(y)=-F(y,0)$.
Letting $t\to \infty $, $\chi (t,s)$ must converge to the equilibrium susceptibility 
whose value is given by the fluctuation dissipation theorem, $\chi _{eq}=T_f^{-1}$.
This leads to the identification of the last term on the r.h.s. of Eq.~(\ref{intermm})
with $\chi _{eq}$~\cite{nota2}, and Eq.~(\ref{intermm}) can be cast as
\be
\chi (t,s)-\chi _{eq} \sim s^{-a_\chi }f(y),
\label{lastintr}
\ee
where $a_\chi =(d-2+\eta )/z$.
Notice that, differently from the case of quenches in the ordered phase, here
the exponent $a_\chi $ is directly related to the equilibrium critical exponents
$\eta $ and $z$.
We recall that, in the KT phase, the exponents $\eta,z,\lambda$ 
depend on temperature. 

It is interesting to discuss the parametric plot of
$\chi (t,s)$ against $C(t,s)$. Since $C(t,s)$ is a monotonically 
decreasing function of $t$, this time can be re-parametrized
in terms of $C$, obtaining $\chi (t,s)=\widehat \chi (C,s)$.
This quantity is important because, if appropriate conditions are
satisfied, its large-$s$ limit
\be
\widehat \chi (C)=\lim _{s\to \infty}\widehat \chi (C,s)
\ee
provides a connection between static and dynamic properties \cite{Franz98}
through the relation
\be
P(q)=-T_f \left . \frac{d^2\widehat \chi (C)}{dC^2}\right \vert _{C=q},
\ee
where $P(q)$ is the overlap probability function of the equilibrium state
at $T=T_f$. As discussed in~\cite{Corberiteff}, a universal linear relation
\be
T_f\widehat \chi (C)=T_f\chi _{eq}-C
\label{parametric}
\ee
as for equilibrated systems,
is expected for quenches to a critical point or into a critical phase, 
although the system is aging for any finite time. 

\subsection{Numerical results}\label{koster_2}
In this section we present results of simulations of the model with $p=6$ 
quenched to $T_1<T_f=0.76<T_2$. 
A typical configuration of domains in the late stage is shown in 
Fig.~\ref{figdomainsp6T076}. In this case, there are
no compact domains.
\begin{figure}
    \centering
    \caption{(Color online) Configuration of the system at $t=3200$ after a quench to $T_f=0.76$.}
\label{figdomainsp6T076}
\vspace{2cm}
\end{figure}

The quantity $I(t)$ is shown in Fig.~\ref{figlengthp6kt}. 
Here one observes that the power law behavior sets in very early.
This implies, through Eq.~(\ref{intkt}),
that also $L(t)$ has a power law growth. The effective exponent has a small
tendency to increase as $t$ gets larger:
We measure an exponent $0.34$
in the decade $10^2-10^3$ and $0.35$ for $t>10^4$. 
The exponents $\eta $ and $z$ are known numerically~\cite{Challa86}. 
At $T_f=0.76$ their measure gives $\eta =0.17$ and $z=2.18$,
yielding $(3-d-\eta )/z= 0.38$. This number is consistent with
the value $0.35$ obtained from $I(t)$ by means of Eqs.~(\ref{lung}), taking also into account that 
the effective exponent we measure is still
increasing at the longest simulated times. 

\begin{figure}
    \centering
    \caption{(Color online) The behavior of $I(t)$. The dashed line is the power law $t^{0.35}$.}
\label{figlengthp6kt}
\vspace{2cm}
\end{figure}

In Fig.~\ref{figgrp6T076} we test the scaling form~(\ref{firstintr}) 
of the equal time correlation function. 
We plot $r^{\eta}G_{ag}(r,t)$ against $x$ for several values of
$t$, where $L(t)$ is computed through Eq.~(\ref{lung}). The data show a
very good collapse on a single master-curve $g(x)$.
Notice that, as expected, Porod's behavior at small $x$ is not observed,
due to the non compact nature of the domains. 

\begin{figure}
    \centering
    \caption{(Color online) Data collapse of $r^{\eta}G(r,t)$ against $x$ for several times 
($t=666, 1097, 1809, 2981, 4915, 8104, 13360, 17155$)}
\label{figgrp6T076}
\vspace{2cm}
\end{figure}

We turn now to consider two time quantities.
In Fig.~\ref{figscalcp6T076} the  
the autocorrelation function is shown.
There is a tendency to a better data
collapse for larger times, implying that the scaling symmetry is still not
exactly obeyed. For the two largest values $s=50$ and $s=100$, however, the
collapse is rather good.

\begin{figure}
    \centering
    \caption{(Color online) Data collapse of $C(t,s)$, for 
     $s=10, 20, 50, 100$ (from bottom to top).}
\label{figscalcp6T076}
\vspace{2cm}
\end{figure}

Let us consider the integrated response function, that is shown in Fig.~\ref{figscalchi2p6T076}. 
Here one observes an analogous situation: the
collapse expected on the basis of Eq.~(\ref{lastintr}) is rather good
for the two largest values of $s$.

\begin{figure}
    \centering
    \caption{(Color online) Data collapse of $\chi (t,s)$ in the quench at $T_f=0.76$, 
     for $s=10,20,50,100$ (bottom up).}
\label{figscalchi2p6T076}
\vspace{2cm}
\end{figure}

In Fig.~\ref{figchidicT076} the parametric plot of $\widehat \chi (C,s)$
is shown. For the largest values of $C$, $\widehat \chi (C,s)$ obeys
Eq.~(\ref{parametric}). As $C$ is decreased the curves flatten and
$T_f\widehat \chi (C,s)$ lies below the asymptotic curve~(\ref{parametric}). 
However,  
in the limit of infinite times $t\to \infty$ , which corresponds to $C\to 0$,
each curve must necessarily obey Eq.~(\ref{parametric}), since
$\chi (t,s)$ approaches the equilibrium value $\chi _{eq}=1/T_f$. Then, moving 
toward $C=0$, at some point the curves become steeper in order to meet
the value $\chi _{eq}$ at $C=0$. Changing $s$, the same qualitative behavior is
observed, but the curve gets higher, slowly approaching the asymptotic
form~(\ref{parametric}) for all values of $C$ in the large $s$ limit. 
This pattern of $\widehat \chi (C,s)$
is analogous to what observed
in the spherical model quenched at the critical point
and is expected in full generality whenever a system is quenched to a critical
point or to a critical phase~\cite{Corberiteff}. 
It must be noticed that the convergence to the trivial form~(\ref{parametric}) is very slow 
because it is regulated by the rather small exponent $\eta (T)$~\cite{Corberiteff}.
Since the exponent $\eta (T)$ at the lower transition temperature is 
expected to behave as $\eta (T_1)\simeq 4/p^2$, the asymptotic behavior  
can be arbitrarily delayed increasing $p$.
The simulations presented in this section have been planned out in order to
show at least a glimpse of this convergence. As discussed in Ref.~\cite{Corberiteff},
previous studies of the KT phase of the XY model, for which the same asymptotic form~(\ref{parametric})
is expected, interpreted a preasymptotic non trivial form of $\widehat \chi (C,s)$
analogous to the one of Fig.~\ref{figchidicT076}
as a reminiscence of the parametric plot of the $d=3$ Edwards-Anderson
model~\cite{Berthier01} or used it to infer the asymptotic value of the fluctuation-dissipation
ratio~\cite{Abriet03}. The simulations presented here clearly show, instead, that this pattern is
preasymptotic and the data are consistent with a 
convergence to the expected trivial limiting form~(\ref{parametric}).

\begin{figure}
    \centering
    \caption{(Color online) Parametric plot of $\chi (t,s)$ vs $C(t,s)$ in the quench at $T_f=0.76$, 
     for $s=10,20,50,100$ (bottom up). The dashed line is the asymptotic curve
     $T_f\chi (C)=T_f\chi _{eq}-C$, Eq.~(\ref{parametric}).}
\label{figchidicT076}
\vspace{2cm}
\end{figure}

\section{Conclusions} \label{concl}

In this paper a rather general numerical investigation 
of the off-equilibrium dynamics 
of the clock model after a temperature quench
has been carried out. We have considered both quenches into the
ordered phase $T_f<T_1$ for systems with $p=3$ and $p=6$ 
and a quench to the critical, Kosterlitz-Thouless phase 
$T_1<T_f<T_2$, for $p=6$. In all these cases we analyzed the behavior
of one-time quantities, such as the equal time correlation function or the characteristic length,
and two-times quantities, such as the integrated response and the autocorrelation
function. 
This study provides a quite general
scenario of the scaling properties of the dynamics and allows
the comparison with the behavior of other well studied 
coarsening systems such as the Ising model or the XY model,
which correspond to the case $p=2$ and $p=\infty$, respectively. 
We find that dynamical scaling is obeyed in all the cases considered.
In the ordered region, the dynamical 
exponents are the same of those of the Ising model. 
While the result
$z=2$ for the clock model was already well known~\cite{Kaski83,Liu93}
the values of the exponents $\lambda $ and $a_\chi $
have been measured for the first time and deserve some considerations. 
These values suggest that the clock model
belongs to the non-equilibrium universality class of the Ising model.
Moreover, finding the same values of the exponents for both $p=3$ and $p=6$ implies
that the system is in the same equilibrium universality class for all $p<\infty $.
Finally, the value $a_\chi =1/4$ fits
with the general phenomenological formula~(\ref{aphenom})
for coarsening systems. This strengthens the idea that 
the non trivial
dimensionality dependence of $a_\chi$ predicted by Eq.~(\ref{aphenom}),
may have a general validity for coarsening dynamics.

It has been recently proposed~\cite{claudio,Henkel2} that, limited to the case of systems 
where topological defects are exclusively domain walls, Eq.~(\ref{aphenom})
may be related to the dynamical roughening of the interfaces.
The fact that the value of $a_\chi $ is the same also in the clock model
implies that the asymptotic contribution of vortices to the response function is
not important, at least at the level of the exponent $a_\chi $, and suggests that
the behavior of the interfaces is the unifying feature that makes systems
with different degeneracy $p$ fall into the same universality class. 

{\bf Acknowledgments} - This work has been partially supported
from MURST through PRIN-2004.



\begin{thebibliography}{99}
\bibitem{nota}
In ref.~\cite{Lippiello05} the relation~(\ref{algochi}) was derived 
making explicit reference for simplicity to scalar spins $\sigma _i=\pm 1$.
However the same derivation applies as well to vector spins.
\bibitem{nota2}
This result can be checked directly. In order to do this
it is sufficient to observe that, due to local equilibrium, 
the fluctuation dissipation theorem must be obeyed
also out of equilibrium for short time separations. Namely, 
by defining $X(t,s)=T_fR(t,s)/[\partial C(t,s)/\partial s]$, for $t\simeq s$
one must have $X(t,s)=1$.
In this regime, using Eqs.~(\ref{secintr},\ref{eerree})
one has $X(t,s)=T_f\tilde f(1)/[(d-2+\eta)\tilde h(1)/z-(t-s+t_0)h'(1)/s]$,
where $h'(y)=dh/dy$. For large $s$ the last term in the denominator can be
neglected. Since $C(t,t)\equiv 1$, from Eq.~(\ref{secintr}) one obtains  
$t_0 ^{-(d-2+\eta )/z}\tilde h (1)=1$. Using this expression to eliminate 
$\tilde h(1)$ in $X(t,s)$ gives
$X(t,s)=T_f\tilde f(1)t_0^{-(d-2+\eta )/z}z/(d-2+\eta )$.  
The condition $X(t,s)=1$ for $t\simeq s$ leads then to the result anticipated 
in Sec.~\ref{kt}, namely
$\tilde f(1)zt_0^{-(d-2+\eta )/z}/(d-2+\eta) =\frac{1}{T_f}=\chi _{eq}$.
\end{thebibliography}
\end{document}